\documentclass[twocolumn,showpacs,preprintnumbers]{revtex4}
\usepackage{amsmath,amssymb}
\usepackage{graphicx}
\usepackage{dcolumn}
\usepackage[vcentermath]{youngtab}
\usepackage[dvips]{epsfig,color}
\usepackage{bm}
\begin{document}
\title{Quark structure of the $X(3872)$ and $\chi_b(3P)$ resonances}
\author{J. Ferretti}\author{G. Galat\`a}
\affiliation{INFN, Sezione di Genova, via Dodecaneso 33, 16146 Genova (Italy)}
\affiliation{Universidad Nacional Aut\'onoma de M\'exico, 04510 M\'exico DF, M\'exico}
\author{E. Santopinto}\thanks{Corresponding author: santopinto@ge.infn.it}
\affiliation{INFN, Sezione di Genova, via Dodecaneso 33, 16146 Genova (Italy)}    
\begin{abstract}
We discuss the nature of the $\chi_b(3P)$ and $X(3872)$ mesons: Are the $\chi_b(3P)$'s standard $b \bar b$ mesons or $b \bar b$ states with a significative continuum component? Is the $X(3872)$ a $c \bar c$ state with continuum coupling effects or a meson-meson molecule?
To do that, we compare quark model and unquenched quark model results for the mass barycenter and splittings of the $\chi_b(3P)$ multiplet. Future and more precise experimental results will discriminate between the two interpretations. In the case of the $X(3872)$, we interpret it a $c \bar c$ core plus higher Fock components due to the coupling to the meson-meson continuum, thus we think that it is compatible with the meson $\chi_{c1}(2P)$, with $J^{PC} = 1^{++}$. The $J^{PC} = 1^{++}$ quantum numbers are in agreement with the experimental results found by the LHCb collaboration. In our view, the $X(3872)$'s mass is lower than the quark model's predictions because of self energy shifts. 
We also provide an estimation of the open charm/bottom strong decay modes of the $X(3872)$ and $\chi_b(3P)$ mesons, such as $X(3872) \rightarrow D \bar D^*$ and $\chi_{b2}(3P) \rightarrow B \bar B$, and radiative transitions.
\end{abstract}
\maketitle

\section{Introduction}
The nature of the $X(3872)$ meson, discovered by Belle in $B$ meson decay \cite{Choi:2003ue} and then confirmed by CDF \cite{Acosta:2003zx}, D0 \cite{Abazov:2004kp} and BABAR \cite{Aubert:2004ns}, is still unknown. 
There are two possible interpretations for this resonance: a weakly bound meson-meson molecule \cite{Hanhart:2007yq,Baru:2011rs,Swanson:2003tb,Fulsom:2007ve,Aceti:2012cb} or a charmonium state \cite{Suzuki:2005ha,Meng:2007cx,Pennington:2007xr,Li:2009zu,Li:2009ad,Danilkin:2010cc,charmonium}, both  with $1^{++}$ quantum numbers. 
Indeed, recently the new results from the LHCb collaboration \cite{Aaij:2013zoa} have ruled out the $2^{-+}$ hypothesis \cite{Burns:2010qq} and therefore only the $1^{++}$ one remains. 
For a summary of theoretical interpretations of the $X(3872)$, see Ref. \cite{Swanson:2006st}.
Thus, one has to make an assumption regarding its quark structure, compatible with $1^{++}$ quantum numbers, to study the properties of the $X(3872)$, such as the decay modes.

The first possibility is to consider the $X(3872)$ as a $\chi_{c1}(2^3P_1)$ charmonium state \cite{Suzuki:2005ha,Meng:2007cx,Pennington:2007xr,Li:2009zu,Li:2009ad,Danilkin:2010cc,charmonium}, with $J^{PC}=1^{++}$.  
Indeed, QM's predictions show that the $\chi_{c1}(2^3P_1)$ is the only state with $1^{++}$ quantum numbers and lying approximately in the same energy region as the $X(3872)$.
The relativized QM \cite{Godfrey:1985xj} predicts this state to have a mass of 3.95 GeV.
Nevertheless, in Ref. \cite{charmonium} we have shown that the uncommon properties of the $X(3872)$ are due to its proximity to the $D \bar D^*$ decay threshold and cannot be easily explained within a standard quark-antiquark picture for mesons. 
Specifically, we have demonstated that $X(3872)$'s mass is lower than quark model's predictions because of self energy shifts (see also Refs. \cite{Pennington:2007xr,Li:2009ad}).
Thus, in our picture the $X(3872)$ is compatible with the meson $\chi_{c1}(2P)$, with $J^{PC} = 1^{++}$, and is thus interpreted as a $c \bar c$ core plus higher Fock components due to the coupling to the meson-meson continuum \cite{charmonium}.

The second possibility is to interpret the $X(3872)$ as a $D \bar D^*$ meson-meson molecule with $1^{++}$ quantum numbers \cite{Hanhart:2007yq,Baru:2011rs,Swanson:2003tb,Fulsom:2007ve,Aceti:2012cb}. 
According to Refs. \cite{Voloshin:1976ap}, the $D \bar D^*$ system with $1^{++}$ quantum numbers can be found by pion exchange and forms a meson-meson molecule. 
More recent molecular model calculations \cite{Tornqvist:2004qy}, including quark exchange kernels for the transitions $D \bar D^* \rightarrow \rho J/\Psi$, $\omega J/\Psi$ in order to predict the $\omega J/\Psi$ decay mode of the $X(3872)$ \cite{Swanson:2003tb}, predict large isospin mixing due to the mass difference between $D^0 \bar D^{*0}$ and $D^+ \bar D^{*-}$.
Nevertheless, in Ref. \cite{Hanhart:2007yq} the authors observe that the one-pion exchange binding mechanism should be taken with greater caution in the $D \bar D^*$ case than in the $NN$ one (see also Refs. \cite{Suzuki:2005ha,Braaten:2007ct,Fleming:2007rp}). 

Other studies proposed a diquark-antidiquark interpretation for the $X(3872)$ \cite{Burns:2010qq,Maiani:2004vq} or different quantum number assignments within a charmonium picture \cite{Eichten:2004uh}.

Another important test for the properties of the $X(3872)$ consists of estimating its strong and radiative decay rates \cite{Suzuki:2005ha,Meng:2007cx,Swanson:2003tb}. 
In Ref. \cite{Meng:2007cx}, the authors re-examine the re-scattering mechanism for the X(3872), which decays to $J/\psi  \rho(\omega)$ through the exchange of $D(^*)$ mesons between intermediate states $D(\bar D)$ and $\bar D^* (D^*)$.
Their results for the ratio $R_{\rho/\omega} \approx 1$, between the decay modes $X(3872) \rightarrow J/\psi  \rho$ and $X(3872) \rightarrow J/\psi \omega$, and for the rate $X(3872) \rightarrow D^0 \bar D^0 \pi^0$, favor a charmonium $c \bar c$ interpretation for the $X(3872)$. 
In Ref. \cite{Suzuki:2005ha}, the author uses semi-quantitative methods to study some properties of the $X(3872)$; he points out that the binding mechanism and the production rates are incompatible with the molecule interpretation.
However these results have been criticized in several works \cite{Zanetti:2011ju,Nielsen:2010ij,Artoisenet:2009wk,Thomas:2008ja,Braaten:2007ct,Swanson:2006st}. 
In particular in Ref. \cite{Artoisenet:2009wk}, the authors point out that the production rates in the molecular interpretation are compatible with Tevatron data once the charm-meson re-scattering effects are taken into account.
In Refs. \cite{Bignamini:2009sk,Bauer:2004bc} the authors observe also prompt production from the CDF collaboration and  discuss whether a meson-meson molecule, with a dimension of a few fm and intrinsic fragility, can be promptly produced.
By contrast, Refs. \cite{Swanson:2003tb,Hanhart:2007yq,Baru:2011rs} suggest a molecular interpretation for the $X(3872)$.

The $\chi_b(3P)$ system was observed for the first time by ATLAS in 2012 \cite{Aad:2011ih} and later confirmed by D0 \cite{Abazov:2012gh}. 
The two collaborations gave an estimation for the mass barycenter of the multiplet, $M_{\chi_b(3P)} = 10.530 \pm 0.005 (\mbox{stat.}) \pm 0.009 (\mbox{syst.})$ GeV \cite{Aad:2011ih} and $M_{\chi_b(3P)} = 10.551 \pm 0.014 (\mbox{stat.}) \pm 0.017 (\mbox{syst.})$ GeV \cite{Abazov:2012gh}. Nevertheless, they did not provide results for the mass splittings between the members of the $\chi_b(3P)$ system. 

It is worthwhile noting that in Ref. \cite{Abazov:2012gh} the authors state "Further analysis is underway to determine whether this structure is due to the $\chi_b(3P)$ system or some exotic bottom-quark state".
Indeed, $\chi_b(3P)$ resonances are close to $B \bar B$, $B \bar B^*$ and $B^* \bar B^*$ decay thresholds, thus their wave functions may include significative continuum components. 
We think that the present experimental data \cite{Aad:2011ih,Abazov:2012gh} cannot exclude this possibility. 
Something similar also happens in the case of the $X(3872)$, since its properties are influenced by the $D \bar D^*$ channel \cite{charmonium,Danilkin:2010cc,Suzuki:2005ha,Meng:2007cx,Pennington:2007xr}.

In this paper, we analyze the interpretation of the $X(3872)$ as a $c \bar c$ state with $J^{PC} = 1^{++}$. 
To do that, we give results for the strong and radiative decays of the $X(3872)$ resonance. 
Our results are compared to those of the meson-meson molecular picture.

We also provide unquenched quark model (UQM) and relativized quark model (QM) results \cite{Ferretti:2013vua} for the mass barycenter of $\chi_b(3P)$ states, and, especially, for the splittings between them, that can be used to discuss this particular problem. Indeed, the magnitude of the splittings between $\chi_b(3P)$ multiplet's states is still unknown.
As discussed in the following, our UQM \cite{bottomonium,Ferretti:2013vua,charmonium,Bijker:2009up,Santopinto:2010zza,Bijker:2012zza} and relativized QM \cite{Ferretti:2013vua,Godfrey:1985xj} results for the mass barycenter of the $\chi_b(3P)$ system are in accordance with the present experimental data \cite{Aad:2011ih,Abazov:2012gh}, the difference between our theoretical and the experimental results being within the error of a quark model calculation, of the order of $30-40$ MeV. Thus, only more precise experimental data for the masses of $\chi_b(3P)$ mesons will make it possible to distinguish between the two different interpretations of these resonances. 
See also Ref. \cite{Dib:2012vw}, where the authors calculate the mass barycenter and mass splittings between the members of the $\chi_b(3P)$ multiplet within different potential models.

As in the case of the $X(3872)$, we also give results for the open bottom strong decays and radiative transitions of the $\chi_b(3P)$ states. 
These results, along with those for the masses of the $\chi_b(3P)$ resonances, can be a useful tool to investigate the properties of these mesons, that lay quite close to the first open bottom decay thresholds.

\section{Results}
\subsection{Calculation of the $c \bar c$ spectrum with self energy corrections}
\label{Bare energy calculation within the relativized constituent quark model}
In this section, we show our results for the spectrum of $c \bar c$ mesons with self energy corrections of Ref. \cite{charmonium}.
In our calculation we fitted 
\begin{equation}
	\label{eqn:self-trascendental}
	M_a = E_a + \Sigma(E_a)  
\end{equation}
to the experimental data \cite{Nakamura:2010zzi}.
Here, the physical mass of a meson, $M_a$, results from the sum of a bare energy term, $E_a$, computed within Godfrey and Isgur's relativized QM of Ref. \cite{Godfrey:1985xj}, and a self energy correction, 
\begin{equation}
	\label{eqn:self-a}
	\Sigma(E_a) = \sum_{BC} \int_0^{\infty} q^2 dq \mbox{ } \frac{\left| \left\langle BC \vec q  \, \ell J \right| T^\dag \left| A \right\rangle \right|^2}{E_a - E_{bc}}  \mbox{ },
\end{equation}
computed within the unquenched quark model formalism of Refs. \cite{charmonium,bottomonium,Santopinto:2010zza}.
The values of Godfrey and Isgur's model parameters are taken from Ref. \cite{charmonium}.

\begin{figure}[htbp]
\begin{center}
\includegraphics[width=8cm]{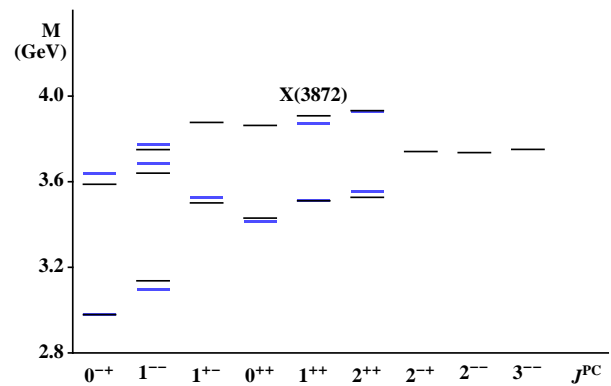}
\end{center}
\caption{Comparison between the calculated masses (black lines) of $1S$, $2S$, $1P$, $2P$ and $1D$ charmonium states via Eq. (\ref{eqn:self-trascendental}) \cite{charmonium} and the experimental ones \cite{Nakamura:2010zzi} (blue boxes). The new values of the parameters of Godfrey and Isgur's model are taken from Ref. \cite{charmonium}. Picture from Ref. \cite{charmonium}; APS copyright.} 
\label{fig:charmonium-spectrum-self}
\end{figure}

According to our result for the mass of the $\chi_{c1}(2^3P_1)$ state ($J^{PC} = 1^{++}$), i.e. 3.908 GeV, the $X(3872)$ is compatible with the meson $\chi_{c1}(2P)$ and includes an extra component due to the coupling to the meson-meson continuum, which is responsible for the downward energy shift. The mass difference between our theoretical prediction and the experimental data is due to the intrinsic error of a QM calculation, of the order of $30-50$ MeV (see Fig. \ref{fig:charmonium-spectrum-self}).

\subsection{Mass barycenter and splittings of the $\chi_b(3P)$ system}
\begin{figure}[htbp]
\begin{center}
\includegraphics[width=5cm]{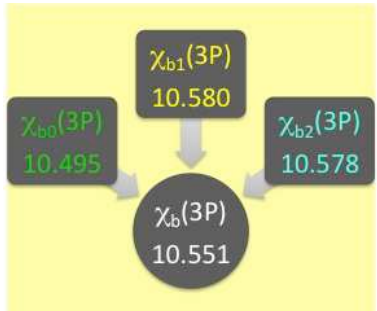}
\end{center}
\caption{Mass barycenter (in GeV) of the $\chi_b(3P)$ system in our UQM calculation, from Ref. \cite{Ferretti:2013vua}.} 
\label{fig:chib3P}
\end{figure}

Here, we show our UQM result for the masses of $\chi_b(3P)$ mesons with self energy corrections of Ref. \cite{Ferretti:2013vua}.
As in the $c \bar c$ case, in our calculation we fitted (\ref{eqn:self-trascendental}) to the existing experimental data for $b \bar b$ mesons \cite{Nakamura:2010zzi,Mizuk:2012pb} and the physical mass of a meson, $M_a$ is given by Eq. (\ref{eqn:self-a}). 
The values of Godfrey and Isgur's model parameters, used to calculate the $E_a$'s, are taken from Ref. \cite{Ferretti:2013vua}. More details can be found in Refs. \cite{charmonium,Ferretti:2013vua}. Our results are reported in Table \ref{tab:mass-bary} and Fig. \ref{fig:chib3P}. 

Then, we show our QM results for $\chi_b(3P)$ mesons, obtained by re-fitting the relativized QM mass formula \cite{Godfrey:1985xj} to the most recent experimental data \cite{Nakamura:2010zzi,Mizuk:2012pb}. In this case, the Fock space is limited to the $q \bar q$ component, while in the UQM case, one also has higher Fock components such as $q \bar q - q \bar q$. These results \cite{Ferretti:2013vua} are reported in Table \ref{tab:mass-bary-GI}.
\begin{table}[htbp]  
\begin{center}  
\begin{tabular}{ccc}
\hline
\hline \\
$M^{\mbox{th}}_{\chi_b(1P)}$ &  $\Delta M_{21}(1P)$ & $\Delta M_{10}(1P)$    \\
9876                                       &  21                            & 30                               \\ \\
$M^{\mbox{th}}_{\chi_b(2P)}$ &  $\Delta M_{21}(2P)$ & $\Delta M_{10}(2P)$    \\
10242                                     &  13                            & 18                               \\ \\
$M^{\mbox{th}}_{\chi_b(3P)}$ &  $\Delta M_{21}(3P)$ & $\Delta M_{10}(3P)$    \\
10551                                     &  -2                            & 85                              \\ \\
\hline
\hline
\end{tabular}
\end{center}
\caption{Mass barycenters of $\chi_b(1P)$, $\chi_b(2P)$ and $\chi_b(3P)$ systems (column 1) and mass splittings between the members of the $\chi_b(1P)$, $\chi_b(2P)$ and $\chi_b(3P)$ multiplets (column 2 and 3), from our UQM calculation of Ref. \cite{Ferretti:2013vua}. The results are expressed in MeV.}
\label{tab:mass-bary}
\end{table}
\begin{table}[htbp]  
\begin{center}
\begin{tabular}{ccc}
\hline
\hline \\
$M_{\chi_b(1P)}$ & $\Delta M_{21}(1P)$ & $\Delta M_{10}(1P)$    \\
9894                   & 21                            & 30                               \\ \\
$M_{\chi_b(2P)}$ & $\Delta M_{21}(2P)$ & $\Delta M_{10}(2P)$    \\
10241                 & 21                            & 14                               \\ \\
$M_{\chi_b(3P)}$ & $\Delta M_{21}(1P)$ & $\Delta M_{10}(1P)$    \\
10510                 & 17                            & 13                               \\ \\
\hline
\hline
\end{tabular}
\end{center}
\caption{Mass barycenters of $\chi_b(1P)$, $\chi_b(2P)$ and $\chi_b(3P)$ systems (column 1) and mass splittings between the members of the $\chi_b(1P)$, $\chi_b(2P)$ and $\chi_b(3P)$ multiplets (column 2 and 3), from our re-fit of Godfrey and Isgur's mass formula of Ref. \cite{Ferretti:2013vua}. The results are expressed in MeV.}
\label{tab:mass-bary-GI}
\end{table}

\subsection{Calculation of $X(3872) \rightarrow D^0 \bar D^{0*}$ and of the open bottom strong decays of $\chi_b(3P)$ states in the $^3P_0$ pair-creation model}
\label{Strong decay widths} 
Here, we discuss our results for the open charm $X(3872) \rightarrow D^0 \bar D^{0*}$ strong decay and the open bottom strong decay widths of $\chi_b(3P)$ states \cite{Ferretti:2013vua}. 
The decay widths are computed as \cite{charmonium,bottomonium,Ferretti:2013vua,3P0,Ackleh:1996yt,Barnes:2005pb} 
\begin{subequations}
\label{eqn:decay-3P0}
\begin{equation}
	\Gamma_{A \rightarrow BC} = \Phi_{A \rightarrow BC}(q_0) \sum_{\ell, J} 
	\left| \left\langle BC \vec q_0  \, \ell J \right| T^\dag \left| A \right\rangle \right|^2 \mbox{ },
\end{equation}
where $\Phi_{A \rightarrow BC}(q_0)$ is the standard relativistic phase space factor \cite{charmonium,bottomonium,Ferretti:2013vua,3P0,Ackleh:1996yt,Barnes:2005pb}, 
\begin{equation}
	\label{eqn:relPSF}
	\Phi_{A \rightarrow BC} = 2 \pi q_0 \frac{E_b(q_0) E_c(q_0)}{M_a}  \mbox{ },
\end{equation}
\end{subequations}
that depends on the relative momentum $q_0$ between $B$ and $C$ and on the energies of the two intermediate state mesons, $E_b = \sqrt{M_b^2 + q_0^2}$ and $E_c = \sqrt{M_c^2 + q_0^2}$. 

The decay $X(3872) \rightarrow D^0 \bar D^{0*}$ is calculated  as
\begin{equation}
	\label{eqn:XintoDD*}
	\begin{array}{rcl}
	\Gamma_{X \rightarrow D^0 \bar D^{0*}} & = & 2 \Phi_{X \rightarrow D^0 \bar D^{0*}}(q_0) \\ 
	& & \sum_{\ell, J} \left| \left\langle D^0 \bar D^{0*} \vec q_0  \, \ell J \right| T^\dag \left| X \right\rangle \right|^2 \mbox{ },
	\end{array}
\end{equation}
where the factor of 2 is introduced because the $X(3872)$ decays into $D^0 \bar D^{0*}$ or $\bar D^0 D^{0*}$.
The masses of the mesons, involved in the calculation of Eq. (\ref{eqn:XintoDD*}), are chosen as: $M_{D^0} = M_{\bar D^0} = 1864.86$ MeV \cite{Nakamura:2010zzi}, $M_{D^{0*}} = M_{\bar D^{0*}} = 2006.98$ MeV \cite{Nakamura:2010zzi}. Since the mass of the $X(3872)$ is very close to the $D^0 \bar D^{0*}$ threshold, even very small differences can substantially modify the results. Thus, we decided to do the calculations using the two latest measures reported by the PDG \cite{Nakamura:2010zzi}: $M_X(3872) = 3871.85$ MeV \cite{Choi:2011fc} and $M_X(3872) = 3871.95$ \cite{Aaij:2011sn}.
The operator $T^\dag$ inside the $^{3}P_0$ amplitude $\left\langle D^0 \bar D^{0*} \vec q_0  \, \ell J \right| T^\dag \left| X \right\rangle$ is that of Refs. \cite{charmonium,bottomonium}, which contains also a quark form factor and an effective pair-creation strength $\gamma_0^{\mbox{eff}}$. 

Finally, the result of our calculation, obtained with the values of the $^3P_0$ model's parameters of Ref. \cite{charmonium}, are:
\begin{eqnarray}
	 \Gamma_{X \rightarrow D^0 \bar D^{0*}}  & =  & 2.5 \mbox { MeV} \mbox{ }, \\
	\Gamma_{X \rightarrow D^0 \bar D^{0*}} & = & 8.3 \mbox { MeV} \mbox{ },
\end{eqnarray} 
respectively for the choice of $M_X(3872) = 3871.85$ and $M_X(3872) = 3871.95$.
These results must be compared to the existing experimental data: $\Gamma_{X \rightarrow D^0 \bar D^{0*}}  = 3.9^{+2.8+0.2}_{-1.4-1.1} \mbox { MeV}$ \cite{Nakamura:2010zzi,Adachi:2008sua} and $\Gamma_{X \rightarrow D^0 \bar D^{0*}}  = 3.0^{+1.9}_{-1.4}\pm0.9 \mbox { MeV}$ \cite{Nakamura:2010zzi,Aubert:2007rva}.
On the contrary, the open charm decays into $D^0 \bar D^0$ and $D^+ D^-$ are forbidden by selection rules.

The open bottom strong decay widths of $\chi_b(3P)$ states are calculated via Eqs. (\ref{eqn:decay-3P0}).
For the masses of the $\chi_b(3P)$ states, we use our UQM predictions of Ref. \cite{Ferretti:2013vua}. 
Indeed, our relativized QM predictions for $\chi_b(3P)$'s masses are of the order of $40-50$ MeV lower and, most important, below $B \bar B$ decay threshold (10.558 MeV) \cite{Nakamura:2010zzi}. See Tables \ref{tab:mass-bary} and \ref{tab:mass-bary-GI}.
Thus, the presence/absence of these decay channels in the experimental data can be used to validate the QM or the UQM interpretation for these states. Our results are reported in Table \ref{tab:decays.3P}. 
\begin{table}[htbp]  
\begin{center}
\begin{tabular}{ccc}
\hline
\hline \\
State                        & Mass [MeV] \cite{Ferretti:2013vua} & $\Gamma (^3P_0)$ [MeV]       \\
$\chi_{b0}(3P)$        & 10495                                            & --       \\ 
$\chi_{b1}(3P)$        & 10580                                            & --       \\ 
$\chi_{b2}(3P)$        & 10578                                            & 13      \\  \\
\hline
\hline
\end{tabular}
\end{center}
\caption{$^3P_0$ model predictions for the open bottom strong decays of $\chi_b(3P)$ states into $B \bar B$ mesons. The values of the $^3P_0$ model parameters and the UQM predictions for the masses of $\chi_b(3P)$ resonances are taken from Ref. \cite{Ferretti:2013vua}. The results are expressed in MeV. The symbol $-$ means that the corresponding decay is forbidden by selection rules or that it cannot take place because it is below the threshold.}
\label{tab:decays.3P}
\end{table} 

\subsection{QM calculation of $X(3872)$'s and $\chi_b(3P)$'s $E1$ radiative transitions}
In this section, we show our QM results for $X(3872)$'s and $\chi_b(3P)$'s radiative transitions.
The $E1$ radiative transitions of the $X(3872)$ are calculated according to \cite{Barnes:2005pb,Eichten:1978tg,Gupta,Kwong:1988ae}:
\begin{subequations}
\label{eqn:radiative}
\begin{equation}
	\label{eqn:E1X3872}
	\Gamma_{E1} = \frac{4}{3} C_{fi} \delta_{SS'} e_c^2 \alpha \left| \left\langle \psi_f \right| r \left| \psi_i \right\rangle \right|^2 E_\gamma^3 
	\frac{E_f^{(c \bar c)}}{M_i^{(c \bar c)}}  \mbox{ }.
\end{equation}
Here, $e_c = \frac{2}{3}$ is the $c$-quark charge in units of $e$, $\alpha$ is the fine structure constant, $E_\gamma$ is the final photon energy, $E_f^{(c \bar c)}$ is the total energy of the final $c \bar c$ state, $M_i^{(c \bar c)}$ is the mass of the initial $c \bar c$ state, the matrix element 
\begin{equation}
	\left\langle \psi_f \right| r \left| \psi_i \right\rangle = \int_0^\infty r^2 dr \mbox{ } \psi^*_{n_f,L_f}(r) \mbox{ } r \mbox{ } \psi_{n_i,L_i}(r)
\end{equation}
 involves the initial and final radial wave functions, and the angular matrix element $C_{fi}$ is given by
\begin{equation}
	C_{fi} = \mbox{max} (L,L') (2J'+1) \left\{ \begin{array}{rcl} L' & J' & S \\ J & L & 1\end{array} \right\}^2 \mbox{ }.
\end{equation}
\end{subequations}
We calculate the matrix elements of Eqs. (\ref{eqn:radiative}) assuming, for the initial and final states, the wave functions of Godfrey and Isgur's relativized QM \cite{charmonium,Godfrey:1985xj}. 
Finally, our results, obtained with the following values of the masses $M_X(3872) = 3871.85$ MeV \cite{Nakamura:2010zzi,Choi:2011fc}, $M_{J/\Psi} = 3096.92$ MeV \cite{Nakamura:2010zzi}, $M_{\Psi(2S)} = 3686.11$ MeV \cite{Nakamura:2010zzi}, $M_{\Psi(3770)} = 3773.15$ MeV \cite{Nakamura:2010zzi} and $M_{\Psi_{2}(1^{3}D_{2})} = 3838$ MeV \cite{Swanson:2003tb} are reported in Table \ref{tab:e.m.-decays}. 
\begin{table}[htbp]  
\begin{center}
\begin{tabular}{ccccc} 
\hline 
\hline \\
Transition                                                                 & $E_{\gamma}$ & $\Gamma_{c \bar c}$      & $\Gamma_{D \bar D^*}$ & $\Gamma_{exp.}$ \\ 
                                                                                & [MeV]               & [KeV]                               & [KeV]                               &  [KeV]                    \\ \\
\hline \\
$X(3872) \rightarrow J/\Psi \gamma$                      &  697                & 11                                   &      8                                &  $\approx 7$          \\ 
$X(3872) \rightarrow \Psi(2S) \gamma$                   &  181                & 70                                   &     0.03                            &  $\approx 36$         \\ 
$X(3872) \rightarrow \Psi(3770) \gamma$               &  101                & 4.0                                  &      0                                &          \\ 
$X(3872) \rightarrow \Psi_{2}(1^{3}D_{2}) \gamma$ &  34                  & 0.35                                &      0                                &          \\  \\ 
\hline 
\hline
\end{tabular}
\end{center}
\caption{$E1$ radiative transitions of the $X(3872)$, calculated with Eq. (\ref{eqn:radiative}) (see column 3). We assume Godfrey and Isgur's relativized QM wave functions \cite{charmonium,Godfrey:1985xj}. In column 4 are reported the molecular model predictions of Ref. \cite{Swanson:2003tb}. The experimental values of column 5 are only estimated from the upper or lower limits given by the PDG \cite{Nakamura:2010zzi} for each decay.}
\label{tab:e.m.-decays}  
\end{table}

The calculation of radiative and strong decays of the $X(3872)$, including also loop corrections, will be the subject of a subsequent paper.

The radiative transitions of $\chi_b(3P)$ states can be calculated analogously. 
In this case, Eq. (\ref{eqn:E1X3872}) can be re-written as \cite{Barnes:2005pb,Eichten:1978tg,Gupta,Kwong:1988ae}:
\begin{equation}
	\Gamma_{E1} = \frac{4}{3} C_{fi} \delta_{SS'} e_b^2 \alpha \left| \left\langle \psi_f \right| r \left| \psi_i \right\rangle \right|^2 E_\gamma^3 
	\frac{E_f^{(b \bar b)}}{M_i^{(b \bar b)}}  \mbox{ }.
\end{equation}
where, $e_b = -\frac{1}{3}$ is the $b$-quark charge in units of $e$, $E_f^{(b \bar b)}$ is the total energy of the final $b \bar b$ state and $M_i^{(b \bar b)}$ is the mass of the initial $b \bar b$ state. 
We calculate the matrix elements of Eqs. (\ref{eqn:radiative}) assuming, for the initial and final states, the wave functions of Godfrey and Isgur's relativized QM \cite{charmonium,Godfrey:1985xj}. 
Finally, our results, obtained with the values of the masses of Table \ref{tab:masses}, are reported in Table \ref{tab:e.m.-decays-chiB}. 
\begin{table*}
\begin{tabular}{cccc}
\hline
\hline \\
State                            & Mass(QM) [MeV] & Mass(UQM) [MeV] & Mass(Exp) [MeV]     \\ \\
\hline \\
$\Upsilon(1^3S_1)$     &  9490                 & 9489                   & 9460   \\
$\Upsilon(2^3S_1)$     &  10012               & 10022                 & 10023 \\
$\Upsilon(1^3D_1)$    &  10144               & 10112                 & --       \\
$\Upsilon_2(1^3D_2)$ &  10154               & 10121                & 10164   \\
$\Upsilon_3(1^3D_3)$ &  10161               & 10127                & --   \\
$\Upsilon(3^3S_1)$     &  10338               & 10358                 & 10355 \\
$\Upsilon(2^3D_1)$    &  10425               & --                       & --       \\
$\Upsilon_2(2^3D_2)$ &  10432               & --                      & --   \\
$\Upsilon_3(2^3D_3)$ &  10438               & --                      & --   \\
$\chi_{b0}(3P)$            & 10494                & 10495                & --       \\ 
$\chi_{b1}(3P)$            & 10511                & 10580                & --       \\ 
$\chi_{b2}(3P)$            & 10524                & 10578                & --      \\  \\
\hline
\hline
\end{tabular}
\caption{Mass of the $b \bar b$ mesons used in the calculation of Eq. (\ref{eqn:radiative}). The theoretical values, resulting from a QM and an UQM calculation, are extracted from Ref. \cite{Ferretti:2013vua}. When available, we use the experimental value from the PDG \cite{Nakamura:2010zzi}.}
\label{tab:masses}
\end{table*}

\begin{table*}
\begin{tabular}{ccccc} 
\hline 
\hline \\
Transition                                                                                 & $E_{\gamma}(\mbox{QM})$ & $\Gamma_{b \bar b}(\mbox{QM})$ & $E_{\gamma}(\mbox{UQM})$ & 
            $\Gamma_{b \bar b}(\mbox{UQM})$  \\    
                                                                                                & [MeV]               & [KeV]        & [MeV]                           &    [KeV] \\ \\
\hline \\       
$\chi_{b0}(3^3P_0) \rightarrow \Upsilon(1^3S_1) \gamma$     & 983                &   0.6         & 984                               & 0.2 \\ 
$\chi_{b0}(3^3P_0) \rightarrow \Upsilon(2^3S_1) \gamma$     & 460                &   1.2         & 461                               & 0.7 \\
$\chi_{b0}(3^3P_0) \rightarrow \Upsilon(3^3S_1) \gamma$     & 138                &    6.1        & 139                               & 6.4 \\
$\chi_{b0}(3^3P_0) \rightarrow \Upsilon(1^3D_1) \gamma$     & 344                &   0.2        & 376                               & 0.2 \\ 
$\chi_{b0}(3^3P_0) \rightarrow \Upsilon(2^3D_1) \gamma$     & 69                 &   0.9         & --                                  & --  \\ \\                                                                                         
$\chi_{b1}(3^3P_1) \rightarrow \Upsilon(1^3S_1) \gamma$     & 971                 &   2.1         & 1061                            & 1.7 \\ 
$\chi_{b1}(3^3P_1) \rightarrow \Upsilon(2^3S_1) \gamma$     & 477                 &   2.5       & 542                              & 2.7 \\
$\chi_{b1}(3^3P_1) \rightarrow \Upsilon(3^3S_1) \gamma$     & 155                 &    7.4         &  223                             & 22.8 \\
$\chi_{b1}(3^3P_1) \rightarrow \Upsilon(1^3D_1) \gamma$    & 361                 &     0          & 458                              & 0 \\
$\chi_{b1}(3^3P_1) \rightarrow \Upsilon(2^3D_1) \gamma$    & 86                   &     0.4       & --                                & -- \\
$\chi_{b1}(3^3P_1) \rightarrow \Upsilon_2(1^3D_2) \gamma$ & 341                &      0         & 408                              & 0.1  \\
$\chi_{b1}(3^3P_1) \rightarrow \Upsilon_2(2^3D_2) \gamma$ & 79                  &     1.0       & --                                 & -- \\  \\
$\chi_{b2}(3^3P_2) \rightarrow \Upsilon(1^3S_1) \gamma$     & 1010              &   3.9         & 1059                             & 3.6 \\ 
$\chi_{b2}(3^3P_2) \rightarrow \Upsilon(2^3S_1) \gamma$     & 489                &   3.8         & 540                               & 4.0 \\
$\chi_{b2}(3^3P_2) \rightarrow \Upsilon(3^3S_1) \gamma$     & 168                &   8.2         &  221                              & 19.6 \\
$\chi_{b2}(3^3P_2) \rightarrow \Upsilon(1^3D_1) \gamma$     & 373               &   0            & 456                               & 0 \\ 
$\chi_{b2}(3^3P_2) \rightarrow \Upsilon(2^3D_1) \gamma$     & 99                 &   0            & --                                 & -- \\ 
$\chi_{b2}(3^3P_2) \rightarrow \Upsilon_2(1^3D_2) \gamma$ & 354                &   0           & 406                               & 0 \\ 
$\chi_{b2}(3^3P_2) \rightarrow \Upsilon_2(2^3D_2) \gamma$ & 92                 &   0.3         & --                                  & --\\ 
$\chi_{b2}(3^3P_2) \rightarrow \Upsilon_3(1^3D_3) \gamma$ & 357               &   0            & 441                                & 0 \\ 
$\chi_{b2}(3^3P_2) \rightarrow \Upsilon_3(2^3D_3) \gamma$ & 86                 &   1.4         & --                                  & --\\ \\                                                                                
\hline 
\hline
\end{tabular}
\caption{$E1$ radiative transitions of the $\chi_b(3P)$, calculated with Eq. (\ref{eqn:radiative}). We use the wave functions resulting from a QM or an UQM calculations.}
\label{tab:e.m.-decays-chiB}  
\end{table*}

\section{Conclusion}
In this paper, we analysed some properties of the $X(3872)$, to understand its nature and quark structure: $c \bar c$ state (plus higher Fock components) or $D \bar D^*$ molecule?

The starting point of our analysis was obviously the spectrum.
Specifically, we calculated the spectrum of $c \bar c$ mesons with self energy corrections and argued that, according to our results,  the $X(3872)$ is compatible with the meson $\chi_{c1}(2P)$ and includes an extra component due to the coupling to the meson-meson continuum, which is responsible for the downward energy shift \cite{charmonium}.

The following step was to study the decay modes of the $X(3872)$ and, in particular, the strong and electromagnetic ones.
Indeed, the radiative and strong decays of the $X(3872)$ offer a promising method to distinguish the charmonium and molecular assignments for this state \cite{Swanson:2003tb}. 
Our prediction for the $X(3872) \rightarrow D^0 \bar D^{0*}$ decay width, resulting from a $^3P_0$ model \cite{Ackleh:1996yt,charmonium,Barnes:2005pb,3P0} calculation, is compatible with the present (preliminary) experimental data \cite{Nakamura:2010zzi,Adachi:2008sua,Aubert:2007rva} and seems to favour a charmonium interpretation for the $X(3872)$.

The results for $E1$ radiative transitions of the $X(3872)$ seem to be quite sensitive to different model details and interpretation for the meson, i.e. charmonium state or meson meson molecule. 
Specifically, when one compares QM to molecular model's calculations, as it is done in Table \ref{tab:e.m.-decays}, it is clear that the two different pictures provide completely different results.
In particular, the absence of the two decay modes $X(3872) \rightarrow \Psi(3770) \gamma$ and $X(3872) \rightarrow \Psi_{2}(1^{3}D_{2}) \gamma$ from molecular model's predictions could be an important clue to distinguish between the two theoretical interpretation for the $X(3872)$.

We also discussed the nature of the $\chi_b(3P)$ mesons: standard $b \bar b$ mesons or $b \bar b$ mesons with strong threshold effects? 
Our QM and UQM results for the mass barycenters and mass splittings for these states, when compared to the future experimental data, will allow to distinguish between the two interpretations. 
In particular, it will be shown if our UQM predictions are compatible with the experimental data or if the difference between them and the QM's are only due the difficulties of fitting a large quantity of states in the $b \bar b$ sector within the UQM. 
This also holds for our calculations of the open bottom strong decays and radiative transitions of the $\chi_b(3P)$ resonances. 
Indeed, it is worthwhile noting that the decay into $B \bar B$ mesons is only compatible with the UQM prediction for the masses of the $\chi_b(3P)$'s. 

\begin{acknowledgments}
We acknowledge useful discussions with P. Colangelo.
This work was supported in part by INFN and in part by CONACYT (Grant No. 78833) and by PAPIIT-DGAPA (Grant No. IN113711), Mexico.
\end{acknowledgments}


\end{document}